\documentclass[12pt,preprint]{aastex}

\begin{document}

\title {Externally-Polluted White Dwarfs With Dust Disks }

\author{M. Jura\footnote{Department of Physics and Astronomy and Center for Astrobiology, University of California, Los Angeles CA 90095-1562; jura, ben @astro.ucla.edu}, J. Farihi,\footnote{Gemini Observatory, 670 North A'ohoku Place, Hilo HI 96720; jfarihi@gemini.edu}\,\,\,\,\&\, B. Zuckerman$^{1}$}

\begin{abstract}
We report {\it Spitzer Space Telescope}  photometry of eleven externally-polluted white dwarfs.  Of the nine stars for which we have IRAC photometry,  we find that GD 40,  GD 133 and PG 1015+161 each has an infrared excess that  can be understood as arising from a flat, 
 opaque, dusty disk.   GD 56 also has an infrared excess characteristic of circumstellar dust, but a flat-disk model  cannot reproduce the data unless there are grains as warm as 1700 K and perhaps not even then.   Our data  support the previous suggestion that
  the metals in the atmosphere of GD 40  are the result of accretion  of a tidally-disrupted asteroid with a chondritic composition.   
\end{abstract}

\section{INTRODUCTION}

Gravitational settling of heavy elements is so effective in white dwarfs cooler than 20,000 K that their atmospheres are expected to be pure hydrogen or pure helium (Paquette et al. 1986), and the ${\sim}$20\% of these stars which
exhibit photospheric metals are thought to be externally polluted (Wolff et al. 2002, Zuckerman et al. 2003,  Koester \& Wilkin 2006). A promising model to explain these data is that a minor-body is perturbed (Debes \& Sigurdsson 2002) to orbit within the tidal radius of the star where it is destroyed and a dusty disk is produced (Jura 2003).  Accretion from this disk can explain atmospheric pollutions (Zuckerman et al. 2003) as seems likely to have occurred for  GD 362 (Becklin et al. 2005, Kilic et al. 2005, Jura et al. 2007) and G29-38 (Jura 2003, Reach et al. 2005b).  While the disrupted minor-body model is promising, despite extensive surveys, only a handful of white dwarfs are known to display an infrared excess (Kilic et al. 2006b, Kilic \& 
Redfield 2007, Mullally et al. 2007, Farihi et al. 2007).  We have therefore obtained {\it Spitzer Space Telescope} data to investigate the frequency of dusty disks orbiting white dwarfs and to assess models for the sources of the external pollution.

\section{OBSERVATIONS}

We observed eleven stars with atmospheric metals  found in   
the SPY project (Koester et al. 2005) that also show  a hint of K-band excess in  2MASS photometry.  
 We
 used  the IRAC (Fazio et al. 2004) and MIPS (Rieke et al. 2004)
cameras on the {\it Spitzer Space Telescope} (Werner et al. 2004) to obtain photometry at wavelengths between 3.6 ${\mu}$m and 24 ${\mu}$m.   IRAC observations were executed using a 20 point cycling
dither pattern of medium step size, with 30 s individual exposures,
yielding a total integration time of 600 s at all wavelengths.  MIPS
observations were performed using 10 cycles of the default 14 point
dither pattern, with 10 s individual exposures, yielding a total
integration time of 1400 s.  The data were processed with the IRAC
and MIPS calibration pipelines (both versions 14) to create single,
reduced images upon which to make photometric measurements.  We removed the one dimensional artificial 
gradient in the sky which is a common feature of MIPS pipelined data by subtracting a median
collapsed image across the direction of the gradient.  Aperture 
photometry was carried out with standard IRAF tasks, and measured
fluxes were corrected for aperture size, but not for color. For IRAC detections, fluxes were typically measured in a 3-pixel radius aperture, but a 2-pixel radius aperture was used for faint  sources or objects in a crowded field, and a 5-pixel radius aperture was used for GD 56, the brightest source.  MIPS 24 ${\mu}$m flux was measured in a 2.45 pixel radius aperture for GD 56, but a smaller aperture with a radius of 1.22 pixels was necessary for GD 133 because of a nearby galaxy.  The
measured fluxes were corrected to the standard IRAC and MIPS 
photometric apertures using corrections found in the most
recent versions of the IRAC and MIPS data handbooks.  The results are listed in Table 1.  IRAC data were not obtained for two sources since they are in other {\it Spitzer} programs; these data have not yet been published.

For both the IRAC and MIPS data, the photometric errors were estimated by taking the per pixel standard deviation in the extracted sky level and multiplying by the area of an aperture with a radius of 2 pixels.  Color corrections have been ignored and are typically less than 1\%.  Our dither pattern should remove the pixel phase dependent correction as well as the IRAC array location dependent errors described by Reach et al. (2005a).  Total 1${\sigma}$ errors (5\% calibration plus photometric measurement errors) are listed in Table 1.  Because there is a somewhat larger scatter in the ratios of  even the best-measured photospheric fluxes in the IRAC bands for white dwarfs (Farihi et al. 2007) than expected from model atmospheres (Tremblay \& Bergeron 2007 and Bergeron 2007, private communication), to be conservative, following Hines et al. (2006) and Silverstone et al. (2006),
we adopt a 10\% calibration uncertainty in the reported fluxes, as reflected in the error bars displayed in Figures 1-4.  Upper limits for non-detections in both IRAC and MIPS images were derived identically to the photometric errors for detections.  

To assess the reliability of our IRAC photometry obtained with small apertures, we also
reduced our data with CCDCAP\footnote{described at:http://www.noao.edu/noao/staff/mighell/ccdcap}. Although there is no significant difference with apertures of
3 pixel-radii, it was found that IRAF reported fluxes 3\% smaller than CCDCAP with 2-pixel radius apertures at the two shorter wavelength IRAC channels where undersampling of the array is most significant.  We have found that this small offset is largely mitigated by the aperture corrections provided by the {\it Spitzer} Science Center, and the net uncertainty introduced by using a small aperture is  well below our 1${\sigma}$ errors.  As a further check, using 2-radius pixel aperture photometry and IRAF, we have
reproduced the IRAC fluxes reported by Reach et al. (2005b) for G29-38 to within 3\% at 4.5 ${\mu}$m and to within 1\% at 7.9 ${\mu}$m.
  We used 2-pixel radius apertures to reduce the data at 3.6 ${\mu}$m and 4.5 ${\mu}$m for only one star, PG 2354+159.

The data for PG 1015+161 are contaminated by a star  ${\approx}$ 2{\arcsec} away (Kilic et al. 2006b). Although our IRAC images  are slightly elongated by 0{\farcs}5 or 0.4 pixels at 3.6 ${\mu}$m and 4.5 ${\mu}$m in the direction of this background star, we cannot accurately deconvolve the image, and  we adopt an alternative approach to estimate its fluxes.   As shown in Figure 5,  a 1995.9 epoch Keck Observatory NIRC z-band image  (see Farihi, Becklin \& Zuckerman 2005), the angular separation of the background star was appreciably greater in the past, 2{\farcs}9 vs. 1{\farcs}8 in 2007.0.  From the Keck data, we find  values of $z$ = 19.12 mag,  $J$ = 18.76  mag and $K$ = 18.26 mag and  infer a spectral type  of approximately K0V.   We extrapolate  these measures to IRAC wavelengths, and  we list in Table 1 our results for PG 1015+161 after the
estimated contributions from the background star   are subtracted from the total measured IRAC fluxes at the position of PG 1015+161. 

Both GD 133 and PG 1015+161 were suspected by Kilic et al. (2006b) to have an infrared excess on the
basis of their 2 ${\mu}$m spectra; our IRAC data demonstrate that  these two stars indeed have  excesses.
GD 56 was previously shown to have a 2 ${\mu}$m excess (Kilic et al. 2006b); our data significantly extend the wavelength range
over which the excess is measured.  Except for the slight hint in the 2MASS photometry,
there is no previous measurement of an infrared excess for GD 40.

\section{MODELS}

We first try to account for the infrared data with the passive, opaque flat disk model of Jura (2003).  Here,   the disk's potential vertical stratification is ignored, so that at each distance from the central star all dust grains achieve the same temperature as determined by
the balance between radiative heating and cooling.  Although a more sophisticated treatment
is required to account for data such as the strong silicate feature seen in both the infrared spectrum of G29-38 (Reach et al. 2005b) and GD 362 (Jura et al. 2007),  this simple model  accounts for 
  the currently available data.  

  Since all the target stars are warmer than 7000 K, the received photospheric infrared flux
from the star, $F_{\nu}(*)$, is modeled to better than the  2${\sigma}$ measurement uncertainty as a single-temperature blackbody (Kilic et al. 2006a, Tremblay \& Bergeron 2007), $B_{\nu}$, so that
\begin{equation}
F_{\nu}(*)\;=\;\frac{{\pi}R_{*}^{2}}{D^{2}}\,B_{\nu}(T_{*})
\end{equation}
 where the star of radius, $R_{*}$, and effective temperature, $T_{*}$, lies at distance $D$ from the Sun. For a flat, opaque disk  that is passively illuminated by the host star; the expected flux, $F_{\nu}(d)$,
  is (Jura 2003):
 \begin{equation}
F_{\nu}(d)\;=\;12\,{\pi}^{1/3}\,\frac{R_{*}^{2}\,\cos\,i}{D^{2}}\,\left(\frac{2\,k_{B}\,T_{*}}{3\,h\,{\nu}}\right)^{8/3}\,\frac{h\,{\nu}^{3}}{c^{2}}\,{\int}^{x_{max}}_{x_{min}}\frac{x^{5/3}}{e^{x}\,-\,1}\,dx
\end{equation}
where $i$ denotes the inclination angle of the disk and $i$ = 0$^{\circ}$ corresponds to a face-on 
configuration.  Here,  $x$ = $(h{\nu})/(k_{B}T_{disk})$ where $T_{disk}$ denotes the temperature in
the disk.   Since the disk is opaque, we cannot estimate its mass.

 Table 2 lists  stellar parameters adopted  from Friedrich, Jordan \& Koester (2004), Koester et al. (2005), and Koester \& Wilken  (2006) as well as our estimated disk parameters.  The fits to our {\it Spitzer} data and near-infrared photometry  from 2MASS  or, when available, from the more accurate observations obtained at the IRTF by Kilic et al. (2006b),  are shown in Figures 1-4. We  find satisfactory fits for  GD 40, GD 133 and PG 1015+161 with a range of possible model parameters that are listed in Table 2.  The inner dust temperature is constrained by the excess
 at 3.6 ${\mu}$m, but the fit is not unique.  To fit the 4.5 ${\mu}$m and 5.7 ${\mu}$m fluxes, there can be either a relatively broad range of temperatures with a more  edge-on disk,  or a narrower range of temperatures with a more face-on disk.   In order not to over-predict the fluxes at 8 ${\mu}$m and 24 ${\mu}$m, the outer temperature cannot be too low.  As listed in Table 2, using the thermal profile given in Jura (2003), the models lie within the tidal radius which typically is about a factor of 100 greater than the star's radius (Davidsson 1999). 
 
 We see in Figure 1 that GD 40 can be equally well fit with a disk with an inner  temperature of 1200 K that is nearly edge-on or one with an inner  temperature of 1000 K that is more nearly face-on.  If, however,
 the inner temperature is only 800 K, then we cannot match the flux at 3.6 ${\mu}$m.  Figure 3 displays the results for PG 1015+161.  Again, there is little difference between a model which is more nearly edge-on and an inner temperature of 1200 K  and  a model which is more nearly face-on with an inner temperature near 1000 K.  However, a disk with an inner temperature of 800 K fails to produce enough
 flux at 3.6 ${\mu}$m.  For GD 133, as shown in Figure 4, a model with an inner temperature of 1200 K that is viewed nearly edge-on is only somewhat better than a model with an inner temperature of 800 K that is viewed nearly face-on.  
 
  Kilic et al. (2006b) have shown that GD 56 is unusual in having a particularly marked excess at 2 ${\mu}$m, and as seen in Figure 2, our model Agd56 with an inner disk  temperature of 1200 K completely fails.  An alternative possibility with a less than satisfactory fit is that the inner disk temperature is larger than 1200 K, and  we also show in Figure 2  the results for model Bgd56 with an inner disk temperature of 1700 K and an outer disk temperature of
 400 K (see Table 2).   
A better fit can be achieved if, rather than a flat disk, we  assume that the disk is substantially  warped as driven by the central star's luminosity (Pringle 1996) or otherwise puffed up  by the gravitational field of a planet. To model this scenario, we assume a single temperature blackbody at 1000 K, and, as shown in Figure 2, we can fit the data if this material is a disk with an angular radius of 
 1.4 ${\times}$ 10$^{-10}$ radians as seen from the Earth.   
   
   Our data are inconsistent with a simple model of interstellar accretion to explain the atmospheric metals.   Assume interstellar grains are accreted at  rate ${\dot M_{dust}}$ at the Bondi-Hoyle radius, $R_{init}$, which is typically between 1 and 10 AU (Koester \& Wilken 2006) and then drift inwards  because of Poynting-Robertson drag to a final radius, $R_{final}$, where they are destroyed.   In this scenario,  the expected (see Jura 2006) infrared flux from interstellar (IS) accretion, $F_{\nu}(IS)$, is:
  \begin{equation}
F_{\nu}(IS)\;{\approx}\;\left(\frac{1}{2}\ln\left[\frac{R_{init}}{R_{final}}\right]
\frac{{\dot M_{dust}}\,c^{2}}{{\nu}}\right)/\left(4{\pi}D^{2}\right)
\end{equation}
  This expression is valid for observations at frequencies where $h{\nu}/k$ lies between the minimum and
maximum grain temperature. 
To evaluate equation (3), we adopt  dust accretion rates that are 0.01 of the rates  of accretion of interstellar gas given by Koester \& Wilken (2006) and also use their estimated distances.  We adopt this value of the interstellar dust to gas ratio by mass of 0.01 from Zubko et al. (2004).  For simplicity, we assume   grains are destroyed by sublimation between ${\sim}$10$^{-2}$ AU and ${\sim}$10$^{-3}$ AU, the region where an unshielded grain that acts like a blackbody  attains a temperature near 1200 K.  We therefore adopt $R_{init}$ = 1000 $R_{final}$ although the results are insensitive to the exact value of this ratio.    We show in Table 3, the predicted values of F$_{\nu}$(24 ${\mu}$m) and a comparison between the predicted and observed fluxes, F$_{pred}$/F$_{obs}$,  for the hydrogen-rich stars where the dwell time of metals in the atmosphere is typically less than 100 yr (Koester \& Wilken 2006) and therefore accretion is almost certainly ongoing.  In contrast, for the three helium rich stars in our sample (GD 40, G26-31, PG 2354+159), the atmospheric dwell times for metals are closer
to 3 ${\times}$ 10$^{5}$ yr (Paquette et al. 1986), and it is possible that accretion has stopped but there are still metals lingering in the stellar photosphere.  We see that from Table 3 that the expected flux is always at least an order of magnitude larger than observed and thus this simple model of interstellar accretion is unsatisfactory. 
\section{DISCUSSION}

We find that four of nine white dwarfs  with fluxes measured in the IRAC bands  display evidence of circumstellar dust disks. For GD 40, the deficiency of carbon in the accreted material (Wolff et al. 2002)  is naturally understood if an asteroid of at least 10$^{23}$ g with a chondritic composition was tidally-destroyed (Jura 2006), in agreement with some previous qualitative suggestions (Sion et al. 1990, Aannestad et al. 1993).     

G29-38 was the first white dwarf found to have an infrared excess (Zuckerman \& Becklin 1987) from dust while GD 362 was the second (Becklin et al. 2005, Kilic et al. 2005). With the results reported here and other recent studies (Mullally et al. 2007, Kilic et al. 2006b, Kilic \& Redfield 2007, Farihi et al. 2007),  there are now enough white dwarfs with infrared excesses that it is 
possible to begin to  discern some patterns.  First, all the white dwarfs with an infrared excess also
display atmospheric metals, thus it is highly plausible that the stars are accreting from 
reservoirs of circumstellar material.   Second, Kilic et al. (2006b), Farihi et al. (2007) and Kilic \& Redfield (2007) have shown that the stars with relatively high calcium abundances also tend to display an infrared excess.  Third, the stars
with an infrared excess all have effective temperatures greater than ${\sim}$9500 K and white dwarf cooling
ages less than ${\sim}$1 Gyr\footnote{GD 362 previously was thought to be hydrogen-rich and have a relatively low luminosity and therefore a cooling age well in excess of 1 Gyr (Gianninas, Dufour \& Bergeron 2004).  However, the star is now known to be helium-rich and have a much
larger luminosity and correspondingly shorter cooling age (Zuckerman et al. 2007).  G167-8 has an effective temperature of 7400 K, but its infrared excess, if real, is from much cooler dust than that found in the  systems described here  (Farihi et al. 2007). }.  According to the simulations of Debes \& Sigurdsson (2002), most of the orbital perturbations of  asteroids would occur
during the first several hundred million yr of the white dwarf's cooling, consistent with the data.

The scenario of a tidally-disrupted asteroid explains the data for white dwarfs with both a relatively high photospheric calcium abundance and an infrared excess.  However, the source of the atmospheric metals of those white dwarfs without an infrared excess  is uncertain.   One possibility is that  the particle density is sufficiently low that mutual collisions lead to effective dust destruction.
At, for example, the typical tidal radius of a white dwarf of ${\sim}$0.01 AU,  the orbital speed is near 300 km s$^{-1}$, and even small deviations from circular
orbits can lead to mutual collision speeds in excess of 10 km s$^{-1}$ which result in  grain
destruction.    For stars with an infrared excess, the disks may be so dense that the grains act more like a granular fluid and the mutual collision speeds are small.  Thus, the externally-polluted white
dwarfs without an infrared excess may have gas disks as has been found for  SDSS J122859.93+104032.9 (Gaensicke et al. 2006).

 White dwarfs with a relatively low accretion rate tend not to possess an infrared excess.    Scaling the accretion rates of Koester \& Wilken (2006) by 0.01 as described above,  Figure 6 presents a comparison  of  ${\dot M}_{dust}$ vs. effective  temperature for DAZs with IRAC photometry, distinguishing between those with and without excess emission.   
The  correlation between having an infrared excess and ${\dot M_{dust}}$ shown
in Figure 6 is closely related to the result found by Kilic et al. (2006b) that the stars with a
greater metal abundance are the ones with an infrared excess.   Although the numbers are very limited, the DAZs with ${\dot M}_{dust}$ larger than about 3 ${\times}$ 10$^{8}$ g s$^{-1}$ possess an excess while the stars with lower accretion rates do not.  For comparison, the dust production rate in the zodiacal cloud is about 3 ${\times}$ 10$^{6}$ g s$^{-1}$ (Fixsen \& Dwek 2002) and considerably larger around some  other main-sequence stars.  Even in their advanced evolutionary state, some white dwarfs may possess a population of eroding parent bodies.

\section{CONCLUSIONS}

We have found evidence for dusty disks orbiting four  externally-polluted white dwarfs.  For GD 40, the evidence lends support to the hypothesis that  tidal-disruption of a  carbon-deficient asteroid has occurred.

This work has been partly supported by NASA and is based on observations made with the {\it Spitzer Space Telescope} which is operated by the Jet Propulsion Laboratory, California Institute of Technology, for NASA.   We thank P. Bergeron for sending us models of
white dwarf atmospheres, M. Kilic for useful correspondence and T. von Hippel for a helpful
referee's report.

\newpage
\begin{center}
Table 1 -- Measured Infrared Fluxes$^{a}$
\\
\begin{tabular}{llllllll}
\hline
\hline
Star & Other Name& F$_{\nu}$(3.6 ${\mu}$m) & F$_{\nu}$(4.5 ${\mu}$m) & F$_{\nu}$(5.7 ${\mu}$m) & F$_{\nu}$(7.9 ${\mu}$m) & F$_{\nu}$(24 ${\mu}$m)\\
&&  (${\mu}$Jy) & (${\mu}$Jy) & (${\mu}$Jy) & (${\mu}$Jy) & (${\mu}$Jy) \\
 \hline
 WD 0300-013$^{b}$ & GD 40 & 231(12) & 199(10) & 159(15) & 164(17) & $<$90 \\
 WD 0408-041 & GD 56 & 1090(55) & 1212(61) & 1177(61) & 1112(59) &240(80) \\
 WD 1015+161 & PG 1015+161&  212(11) & 177(9) &145(17)  &126(19)& $<$110 \\
 & PG 1015+161$^{c}$ &  197(11) & 169(9) & 140(17) & 123(19) & $<$110\\
 WD 1116+026 &GD 133 & 592(30)& 527(26)& 472(27) & 464(29) & 310(100) \\
 WD 1124-293 & ESO 439-80& &&&& $<$50 \\
 WD 1204-136 & EC 12043-1337 & 158(5) & 99(4) & 72(14) & 32(17) & $<$70 \\
 WD 1225+006 & HE 1225+0038&  361(18) & 223(12) & 144(16) & 79(20) & $<$80 \\
 WD 1315-110 & HE 1315-1105 &210(11) & 131(7) & 90(15) & 30(18) & $<$80 \\
 WD 1457-086 &PG 1457-086&&&&& $<$90 \\
 WD 2144-079$^{b}$ &G26-31 &191(10) & 122(7) & 65(19) & 42(22) & $<$100\\
 WD 2354+159$^{b}$ & PG 2354+159& 73(5) & 44(4) & 27(17) & $<$21 & $<$110 \\
 \hline
 \end{tabular}
 \end{center}
 $^{a}$1${\sigma}$ errors for detections are given in parenthesis; the procedure for deriving  the upper limits is described in  \S2.  
 \\
 $^{b}$He-rich star (Koester et al. 2005)
 \\
 $^{c}$PG 1015+161 fluxes corrected for the background star, as discussed in ${\S}$2

\newpage
\begin{center}
Table 2 -- Stellar and Disk  Properties 
\\
\begin{tabular}{llllrlrrrr}
\hline
\hline
Star & $T_{*}$ & $R_{*}$ &$R_{*}/D$ & model & $\cos\, i$& $T_{max} $ & $T_{min}$& $R_{min}$ & $R_{max}$  \\
  & (K) & (R$_{\odot}$)& (10$^{-12}$) & & & (K) & (K) & ($R_{*}$) & ($R_{*})$ \\
\hline
GD 40 & 15,200$^{a}$ & 0.013$^{b}$&  3.6 & Agd40&0.2 & 1200 & 600 & 18 & 44\\
            &                &      & & Bgd40 &0.4 &1000& 700 & 22  & 36 \\
            &                  &      && Cgd40& 0.8&800& 650 & 30 & 40\\
GD 56 & 14,400$^{a}$ & 0.015$^{c}$ &3.6 & Agd56 &1.0 & 1200 & 300& 16 & 104 \\
            &                &      & & Bgd56 & 1.0 & 1700 & 400& 10 & 71 \\
PG 1015+161 & 19,300$^{a}$ & 0.014$^{c}$& 2.7 & Apg10 &0.3 & 1200 & 800 & 24 & 42 \\
                           &              &        && Bpg10 & 0.6 & 1000 & 800& 31 & 42 \\
                            &             &        && Cpg10 & 1.0 & 800 & 700 & 42 & 50\\
GD 133 & 12,200$^{a}$ & 0.014$^{c}$&7.0 & Agd133 & 0.2 & 1200 & 300& 13 & 83 \\
               &              &        && Bgd133 & 0.4 & 1000 & 600 & 17 & 33\\
               &                 &    & &Cgd133 & 0.8  & 800 &  600& 23 & 33 \\
\hline
\end{tabular}
\end{center}
$^{a}$Koester et al. (2005)
\\
$^{b}$Friedrich et al.  (2004)
\\
$^{c}$derived from Koester \& Wilkin (2006)
\newpage
\begin{center}
Table 3 -- Predicted $F_{\nu}$(24 ${\mu}$m) for Simple Interstellar Accretion followed by Poynting-Robertson drag
\\
\begin{tabular}{llllr}
\hline
\hline
Star & $D$ & $dM_{dust}/dt$ & F$_{\nu}$(24 ${\mu}$m) & F$_{pred}$/F$_{obs}$\\
 & (pc) & (10$^{9}$ g s$^{-1}$) & (mJy) \\
 \hline
 GD 56 & 74 & 0.26 & 9.9 & 41\\
 PG 1015+161 & 95 & 2.0 & 46 & $>$420   \\ 
 GD 133 & 43 & 0.37 & 42 & 140 \\
 ESO 439-80 & 34 & 0.13 & 23 & $>$460 \\
 EC 12043-1337 & 52 & 0.23 & 18 & $>$260\\
 HE 1225+0038 & 29 & 0.0059 & 1.5 & $>$19 \\
 HE 1315-1105 & 40 & 0.022 & 2.9 & $>$36\\
 PG 1457-086 & 117 & 1.5 & 23 & $>$260 \\
\hline
\end{tabular}
\end{center}
Distances are taken from Koester \& Wilkin (2006) while the dust accretion rates are scaled from their values as described in \S3. 

\begin{figure}
\plotone{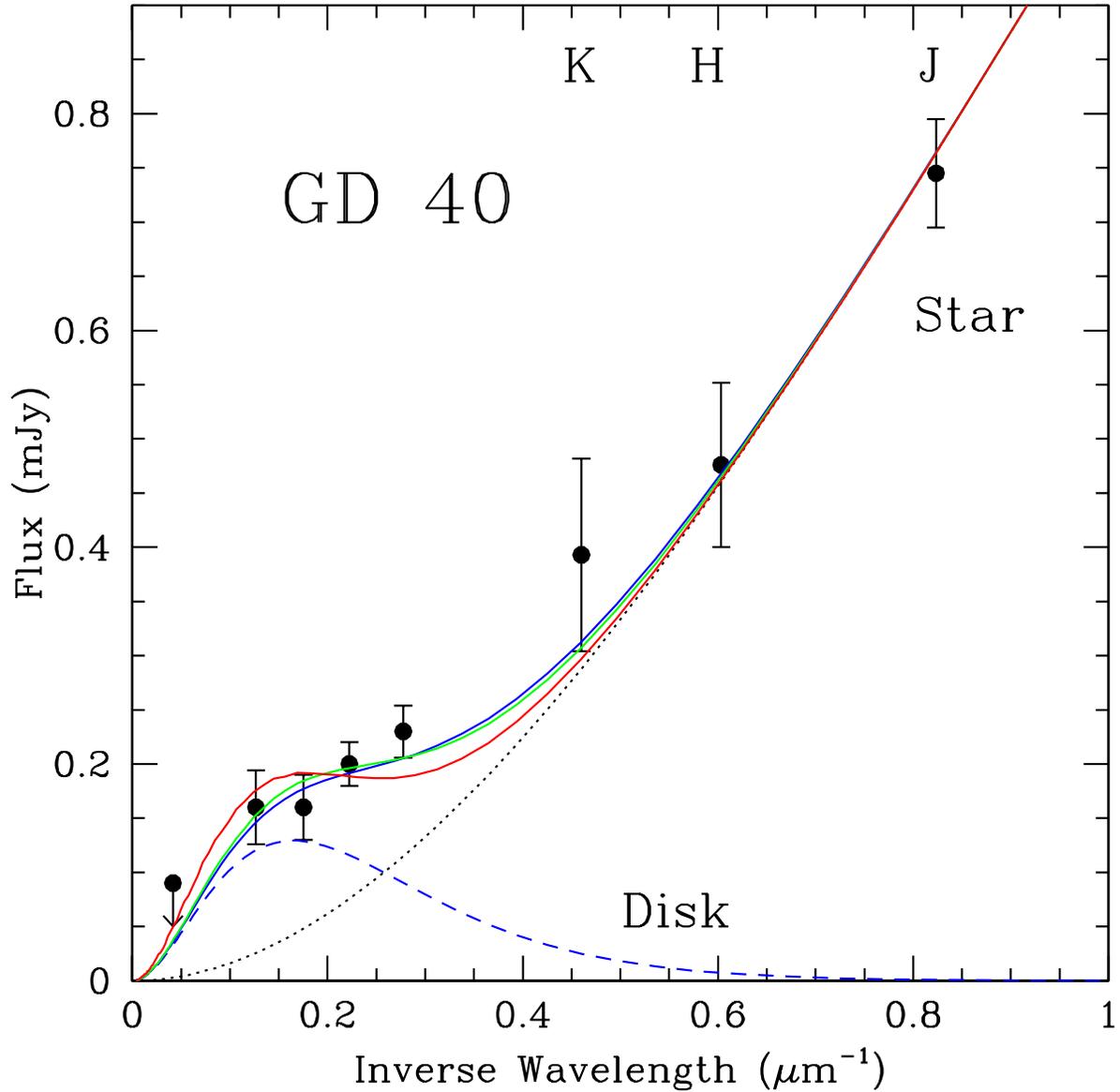}
\caption{Comparison of models from Table 2 and data (2${\sigma}$ errors) for GD 40.  The plot shows the contribution from the stellar photosphere (dotted line), one  disk model (blue dashed line for model Agd40)  and totals (solid blue for model Agd40, solid green for model Bgd40, solid red for model Cgd40).  Models Agd40 and Bgd40 are essentially
indistinguishable while model Cgd40 predicts somewhat too little flux at 3.6 ${\mu}$m because the warmest grains are only 800 K.}
\end{figure}
\begin{figure}
\plotone{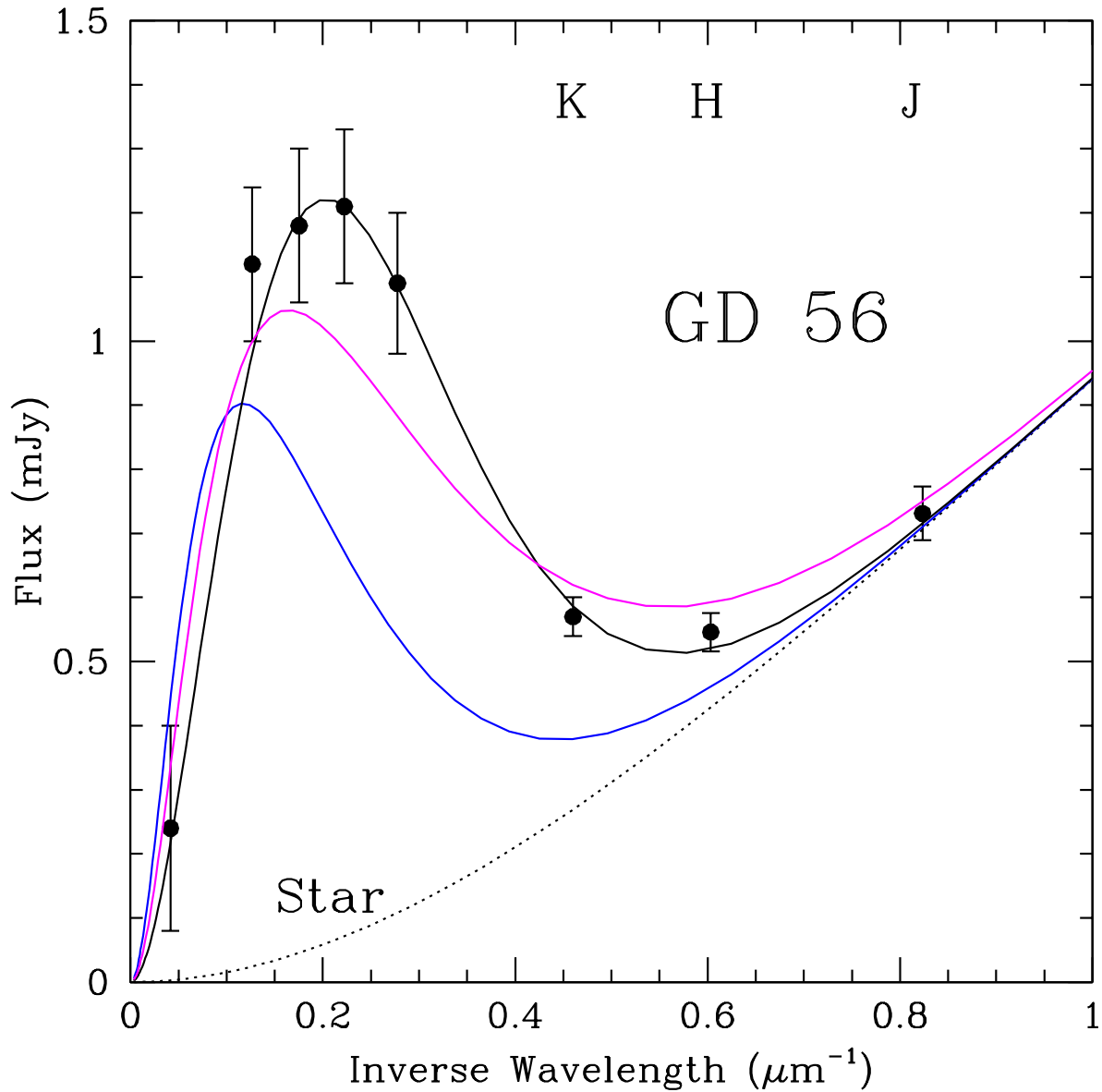}
\caption{Comparison of three models and data (2${\sigma}$ errors) for GD 56.  The dotted line is the contribution from the stellar photosphere.  The sums of the circumstellar and stellar fluxes are shown for  models Agd56 (solid blue line) and Bgd56 (solid magenta line) listed in Table 2.  
We also show (solid black line) the flux for a model summing both the photosphere and  excess emission for an opaque dust cloud at a temperature of 1000 K with an angular radius 1.4 ${\times}$ 10$^{-10}$ radians as seen from  Earth.}
 \end{figure}
 \begin{figure}
 \plotone{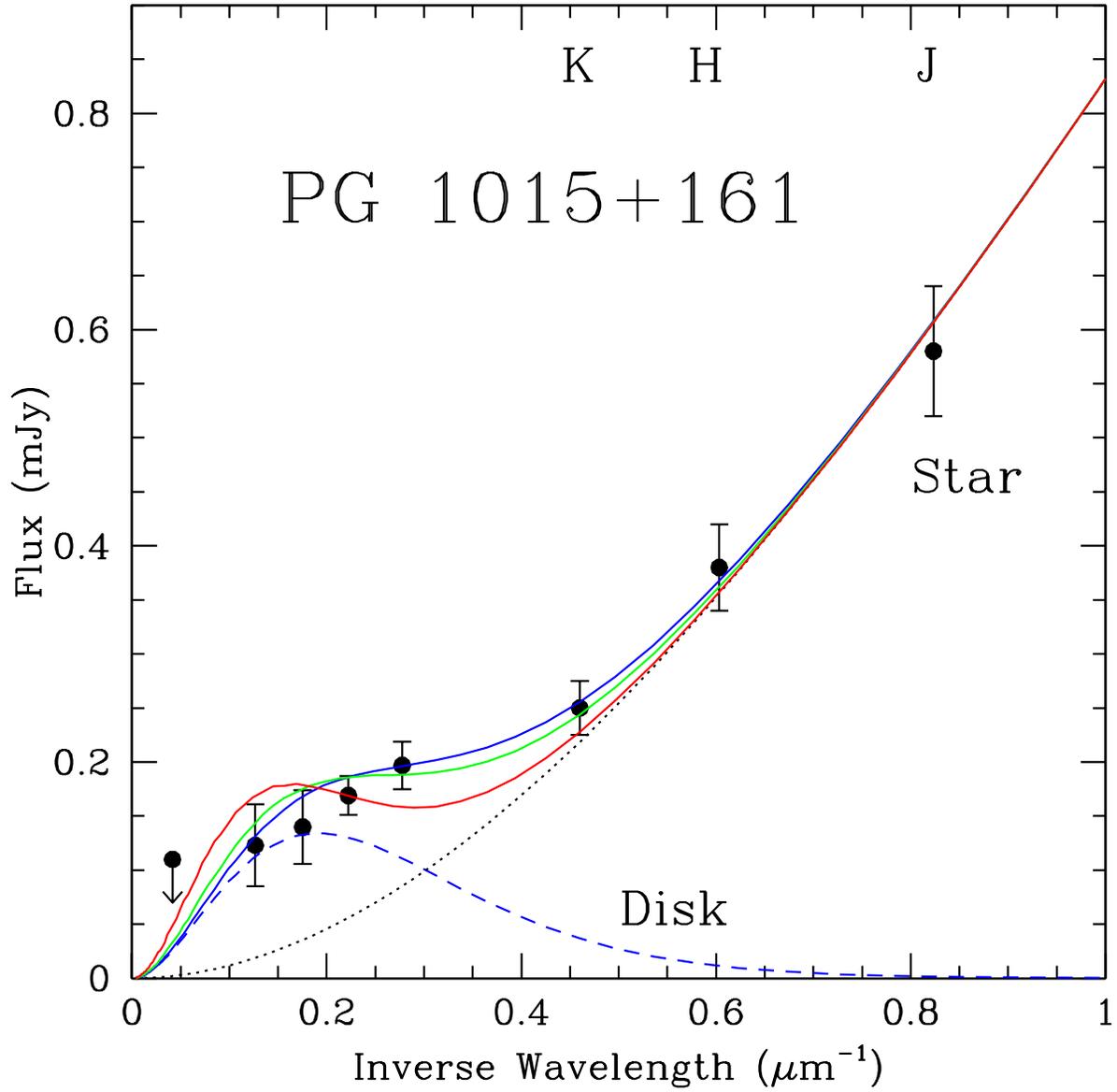}
 \caption{Similar to Figure 1, but for PG 1015+161 with blue, green and red for the totals of models Apg10, Bpg10 and Cpg10, respectively.  The emission from only the disk  is shown for model Apg10}
 \end{figure}
 \begin{figure}
 \plotone{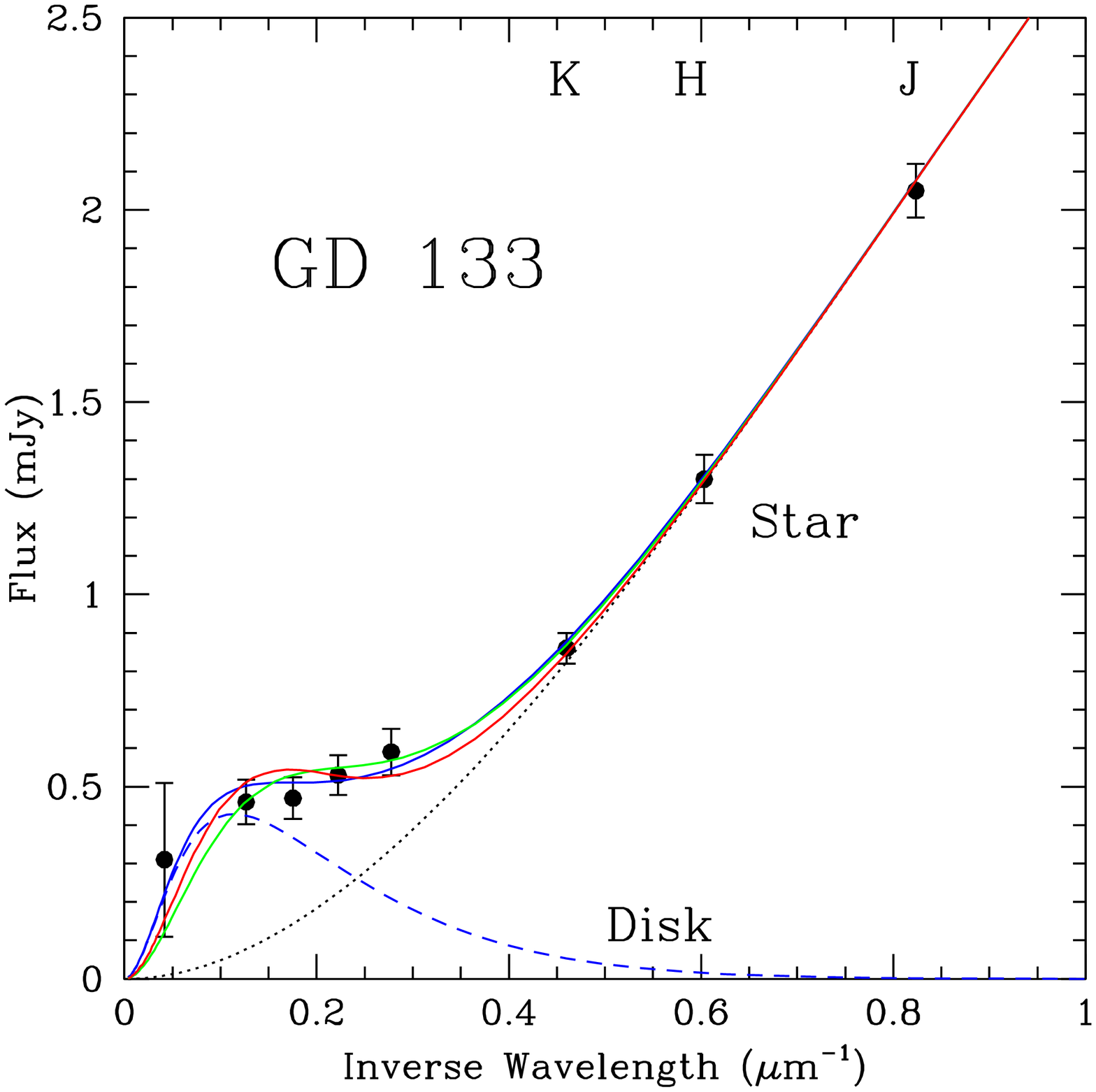}
\caption{Similar to Figure 1, but  for GD 133 with blue, green and red for the totals of models Agd133, Bpg13s and Cpg133, respecitvely.  The emission from only the disk is shown for model Agd133.}
 \end{figure}
 \begin{figure}
\plotone{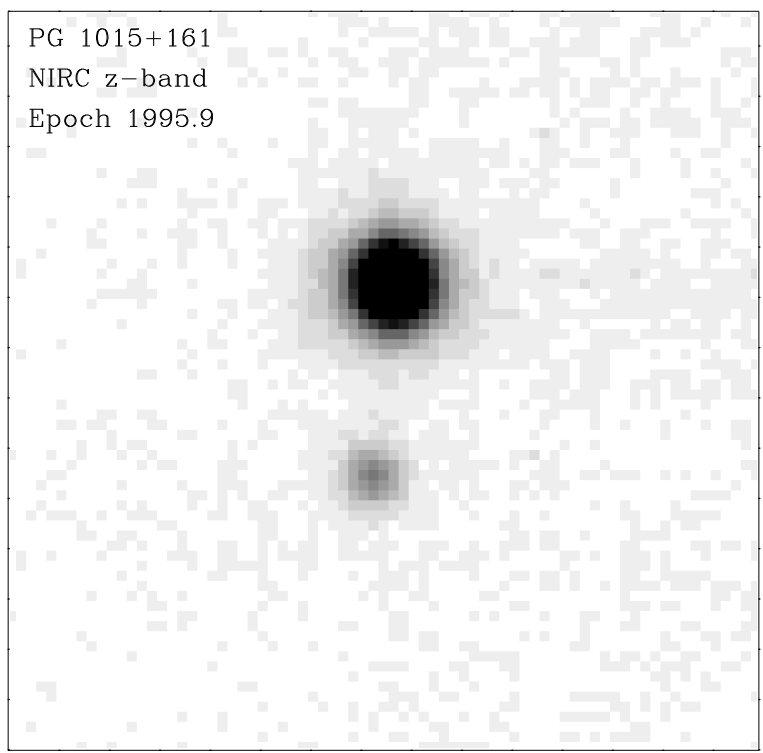}
 \caption{KECK/NIRC z-band image, epoch 1995.9,  obtained of PG 1015+161.  Up lies at P. A. = 96$^{\circ}$; the background star lies  2{\farcs}9 from the white dwarf at  P. A. = 270$^{\circ}$.  The proper motion  proper motion of ${\mu}$ = 0{\farcs}13 yr$^{-1}$ at P. A. =  239$^{\circ}$ (Farihi et al. 2005) of PG 1015+161 is consistent with its  position relative to the background source in our 2007.0 IRAC images.}
\end{figure}
 \begin{figure}
 \plotone{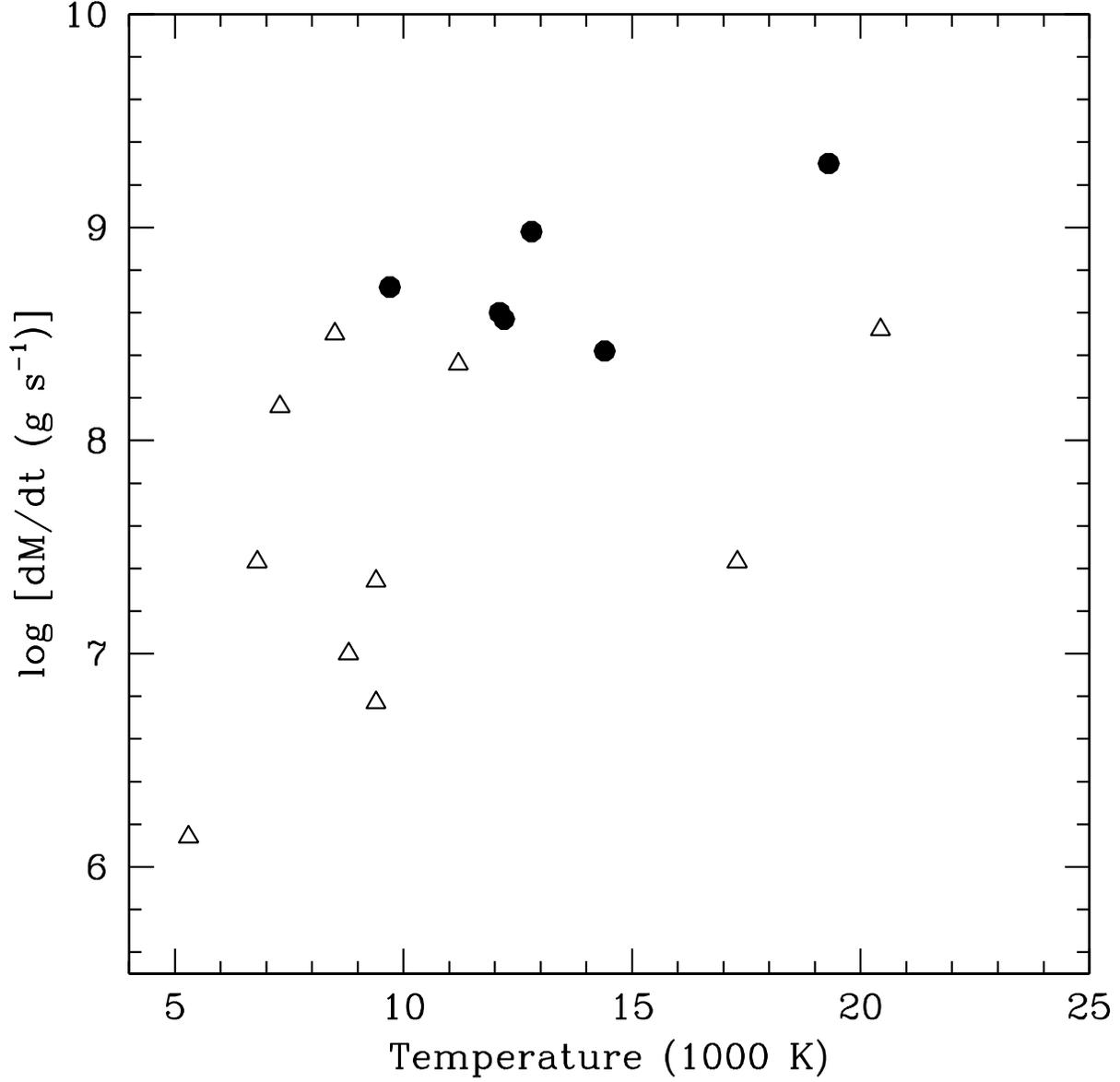}
 \caption{Mass accretion rates (\S3) vs. effective temperature for hydrogen-rich white dwarfs with an excess (solid circles) in either ground-based or ${\it Spitzer}$ data and without  an  excess (open triangles) in published {\it Spitzer} data.   
 GD 133 and  PG 1015+161 (this paper), GD 56 (this paper and Kilic et al. 2006b), G29-38 (Zuckerman \& Becklin 1987), WD 2115-560 (Mullally et al. 2007, von Hippel et al. 2007), WD 1150-153 (Kilic \& Redfield 2007) have  an excess while  EC 12043-1337, HE 1225+038, and HE 1315-1105 (this paper), WD 1202-232, WD 1337+705 and WD 2149+021 (Mullally et al. 2007), and WD 0208+396, WD 0243-026, WD 0245+541, and WD 1257+278 (Debes \& Sigurdsson 2007) do not.  Two helium-rich white dwarfs (GD 40 and GD 362)  that also
 display an excess are not shown here.}
 \end{figure}
\end{document}